\documentclass[11pt]{article}%
\usepackage[iso-8859-7]{inputenc}
\usepackage{epsfig}
\usepackage{graphicx}
\usepackage{epstopdf}
\usepackage{amsfonts}
\usepackage{amssymb}
\usepackage{amsthm}
\usepackage{amsmath}
\usepackage{multirow}%
\providecommand{\U}[1]{\protect\rule{.1in}{.1in}}
\newcommand{\newc}{\newcommand}
\newc{\ra}{\rightarrow}
\newc{\lra}{\leftrightarrow}
\newc{\be}{\begin{equation}}
\newc{\ee}{\end{equation}}
\newc{\ba}{\begin{eqnarray}}
\newc{\ea}{\end{eqnarray}}
\newc{\ov}{\overline}
\newc{\pa}{\partial}
\newc{\D}{\Delta}

\newc{\nn}{\nonumber}

\begin{document}
\thispagestyle{empty}

%\hfill CERN-PH-TH/2012-135
\vskip 2truecm
%%%%%%%%%%%%%%
\vspace*{3cm}

\begin{center}
{ {\bf A discrete anatomy of the neutrino mass matrix}}\\
\vspace*{0.5cm} {G.K. Leontaris$^{\,a}$
and N.D. Vlachos$^{\,b}$}\\
\vspace*{0.5cm} { \normalsize\sl $^a$Theoretical Physics Division,
Ioannina University,
GR-45110 Ioannina, Greece\\[2.5mm]
\normalsize\sl $^b$Theoretical Physics Division, Aristotle
University, GR-54124 Thessaloniki, Greece
 \\[2.5mm]
 }
\end{center}

\vspace*{1cm}

\begin{center}
\textbf{Abstract}
\end{center}

\noindent

We investigate the possibility of expressing the charged leptons
and  neutrino mass matrices as linear combinations of generators of a single
finite group. Constraints imposed on the resulting mixing matrix by current data
restrict  the group types, but allow  a non zero value for the
$\theta_{13}$ mixing angle.

\newpage

\section{On the discrete origin of the neutrino mass}

It is well known that neutrino oscillations are tightly connected to the
existence of non-zero neutrino masses and the mixing in the leptonic sector.
Recent neutrino data are in accordance with two large mixing angles and a tiny
value for the third one. More precisely, adopting the parametrization
\begin{equation}
C= \left(
\begin{array}
[c]{lll}%
c_{12}c_{13} & c_{13}s_{12} & s_{13}\\
-c_{23}s_{12}-c_{12}s_{13}s_{23} & c_{12}c_{23}-s_{12}s_{13}s_{23} &
c_{13}s_{23}\\
s_{12}s_{23}-c_{12}c_{23}s_{13} & -c_{23}s_{12}s_{13}-c_{12}s_{23} & c_{13}c_{23}%
\end{array}
\right) \label{V123}%
\end{equation}
(where $c_{ij}\equiv\cos\theta_{ij}$ etc) for the leptonic mixing matrix, the
$3\sigma$ range of the angles is given by \be
\begin{split}
\sin^{2}\theta_{12} & =[0.26-0.36]\\
\sin^{2}\theta_{23} & =[0.34-0.66]\\
\sin^{2}\theta_{13} & =[0.017-0.031]\cdot
\end{split}
\ee

It has been suggested~\cite{Harrison:2002er}-\cite{Cannoni:2013gq}
that the given structure of the mixing matrix indicates the
existence of underlying symmetries. Such ideas stimulated the
interest on Yukawa superpotentials invariant under discrete groups
such as the permutation groups $S_{4}, A_{4}$ etc. The simplest
implementation of such groups into the lepton sector has predicted a
tri-bimaximal (TB) mixing~\cite{Harrison:2002er} with a strictly
zero value for $\theta_{13}$. However, recent experimental findings
suggest a small non-zero value around $\theta_{13}\approx9^{0}$. In
fact, it has been shown that it  is possible to generate a non-zero
$\theta_{13}$ value even in the context of the minimal TB mixing
scenario by introducing in the neutrino texture
of~\cite{Harrison:2002er} an appropriate
phase~\cite{Xing:2002sw,Leontaris:2011xa} , whose acceptable values
lie within the $\theta_{13}$  experimental range. In the present
work we wish to investigate other possibilities which involve more
elaborate forms of mass textures.  The motivation for such an
analysis is that effective models emerging  from unified theories in
higher dimensions involve a  variety of continuous or discrete
groups which act  as family symmetries. Although these usually imply
mass  matrices which look complicated compared to the one above (see
for example~\cite{Ellis:1998nk}),  yet they are subject to
constraints imposed by the underlying symmetries.

\section{Formulation}

It has been shown that small permutation  groups like $S_{4}, A_{4}$
constitute good approximate  symmetries which, under specific alignments of
the vacuum expectation  values of the various Higgs fields, generate
$m_{\ell,\nu}$ matrices compatible with TB-mixing. Many other finite groups
have also been explored in detail  \cite{Ishimori:2010au}-\cite{King:2013eh}.

One way for obtaining a non-zero value for $\theta_{13}$ is to
consider deformations of the simple  TB mass matrices. Then, we can
explore the possibility of still having an underlying group
structure in this new landscape.   However, since the $\theta_{13}$
value differs  substantially from the values of the other two
angles,  the mixing matrix becomes more involved, hence it is not
obvious whether such a finite symmetry exists.  Consequently, it is
rather difficult to tackle this problem in a straightforward manner.
Therefore, we will follow a different path  and choose to expand the
mass matrices in terms of  appropriate representations of finite
group elements.

\subsection{ Expanding mass matrices in terms of discrete group elements.}

One of the interesting properties of the simple TB mass matrix structure is
that the diagonalising matrix is independent of the mass eigenvalues. In a previous work, having
this in mind and using the Cayley Hamilton theorem ~\cite{Leontaris:2011xa}, we  proposed to
write a $3\times3$ Hermitean mass matrix $M$ in the form
\begin{equation}
M=c_{1}I+c_{2}U+c_{3}U^{2}\label{ExMU}%
\end{equation}
where $U$ is a $3\times3$ unitary matrix. Without loss of generality we
imposed the condition $\det U=1$, so that  $U$ can always be
brought in the form
\begin{equation}
D=\left(
\begin{array}
[c]{ccc}%
e^{i\,\lambda} & 0 & 0\\
0 & 1 & 0\\
0 & 0 & e^{-i\,\lambda}%
\end{array}
\right)  \label{Ueig}%
\end{equation}
by means of a similarity transformation (modulo permutations of the
eigenvalues). Observing that $U$ and $M$ can be simultaneously
diagonalised, it follows that the mixing associated to the mass
matrix $M$ can be simply obtained by diagonalising the matrix $U$.
It is then straightforward to see that the eigenmasses are given
only in terms of the coefficients $c_{i}$, and the value of the
phase $\lambda$, \be \label{cfromm}
\begin{split}
m_{1}-m_{3} &  =2\,i\,\left(  c_{2}\,\sin\lambda+c_{3}\,\sin2\,\lambda\right)
\\
m_{1}+m_{3} &  =2\left(  c_{1}+c_{2}\,\cos\lambda+c_{3}\,\cos2\,\lambda
\right)  \\
m_{2} &  =c_{1}+c_{1}+c_{2} \, \cdot
\end{split}
\ee Hence, the mass eigenvalues problem is essentially disentangled from the
diagonalising matrix.

\section{Deforming the TB-mixing matrix}

 Now, suppose that the
generators $U$ of the mass matrices form elements of a discrete group. It
follows that they must satisfy relations of the form $U^{n}=1$ for some
integer value of $n.$ Their eigenvalues will be $e^{\frac{2\pi i}{n}}$,
$e^{-\frac{2\pi i}{n}}$, $1$ in some order and can be diagonalised by means of
a unitary transformation to produce a diagonal matrix $D_{i,n}$ where the subscript $i$
refers to the eigenvalues ordering. This way, for the leptons we have
\begin{equation}
U_{l}=V_{l}D_{i,n}V_{l}^{\dagger}%
\end{equation}
and for he neutrinos
\begin{equation}
U_{\nu}=V_{\nu}D_{j,m}V_{\nu}^{\dagger}\ .
\end{equation}
Hence, the mixing matrix is
\begin{equation}
C=V_{l}^{\dagger}V_{\nu}\ .
\end{equation}
If $U_{l}$ and $U_{\nu}$ belong to the same group they must satisfy a relation
of the form
\begin{equation}
\left(  U_{l}U_{\nu}\right)  ^{p}=1\ .
\end{equation}%
Also,
\begin{equation}
U_{l}U_{\nu}=V_{l}D_{n}V_{l}^{\dagger}V_{\nu}D_{m}V_{\nu}^{\dagger}=V_{l}%
D_{n}CD_{m}C^{\dagger}V_{l}^{\dagger}=V_{l}\mathcal{T}V_{l}^{\dagger}%
\end{equation}
where
\begin{equation}
\mathcal{T}=D_{n}CD_{m}C^{\dagger}\ \cdot
\end{equation}
This way,%
\begin{equation}
\left(  U_{l}U_{\nu}\right)  ^{p}=V_{l}\mathcal{T}^{p}V_{l}^{\dagger}%
\end{equation}
so we must have $\mathcal{T}^{p}=1$ for some integer $p$. This implies that one
eigenvalue of $\mathcal{T}$ must be $1$ while the other two must be
$e^{\frac{2\pi i}{p}}$, $e^{-\frac{2\pi i}{p}}$respectively, since the
determinant of $\mathcal{T}$ is $1$. The trace of $\mathcal{T}$
equals  $1+2\cos\frac{2\pi}{p}$ . Therefore, we must seek solutions of the form%
\begin{equation}
1+2\cos\frac{2\pi}{p}=\mathrm{Tr}\mathcal{T}\
\end{equation}
for given values of $m$ and $n$. In the cases where $\mathrm{Tr}\mathcal{T}$ takes
complex values there is no solution. Let us now define
\begin{equation}
D_{1,m}=\mathrm{Diagonal}\left[  1,e^{\frac{2\pi i}{m}},e^{-\frac{2\pi i}%
{m}}\right]
\end{equation}%
\begin{equation}
D_{2,m}=\mathrm{Diagonal}\left[  e^{\frac{2\pi i}{m}},1,e^{-\frac{2\pi i}%
{m}}\right]
\end{equation}%
\begin{equation}
D_{3,m}=\mathrm{Diagonal}\left[  e^{\frac{2\pi i}{m}},e^{-\frac{2\pi i}{m}%
},1\right] \cdot
\end{equation}

Next, as a starting point we take the matrix $C$ to be the TB-mixing
matrix. This is readily derived from (\ref{V123}) for
$\sin^{2}\theta_{12}=1/3, \sin^{2}\theta_{23}=1/2, \theta_{13}=0$
and choosing the phases appropriately, $C$ can be written as
\[
C= \left(
\begin{array}
[c]{lll}%
\sqrt{\frac{2}{3}} & -\frac{1}{\sqrt{3}} & 0\\
\frac{1}{\sqrt{6}} & \frac{1}{\sqrt{3}} & -\frac{1}{\sqrt{2}}\\
\frac{1}{\sqrt{6}} & \frac{1}{\sqrt{3}} & \frac{1}{\sqrt{2}}
\end{array}
\right)\, \cdot
\]

A subsequent search leads to a limited number of solutions depicted in Table~\ref{T3}.
The $m$ value which refers to the neutrinos turns out always to be
$2$.

\begin{table}[ptb]
\label{T3}
\par
\begin{center}%
\begin{tabular}
[c]{lllll}%
$D_{i,n}$ & $D_{j,m}$ & $n$ & $m$ & $p$\\
$1$ & $1$ & $3$ & $2$ & $4$\\
$2$ & $2$ & $3$ & $2$ & $3$\\
$3$ & $3$ & $2$ & $2$ & $4$\\
$1$ & $3$ & $n$ & $2$ & $2$\\
$2$ & $3$ & $2$ & $2$ & $4$\\
$3$ & $2$ & $3$ & $2$ & $3$\\
$1$ & $2$ & $3$ & $2$ & $3$%
\end{tabular}
\end{center}
\caption{Solutions for TB-mixing}%
\end{table}

The lepton mass matrix is then written as
\begin{equation}
M_{l}=c_{1}I+c_{2}U_{l}+c_{3}U_{l}^{2}%
\end{equation}%
\begin{equation}
U_{l}=V_{l}D_{i,n}V_{l}^{\dag}%
\end{equation}
and the neutrino mass matrix is written as
\begin{equation}
M_{\nu}=d_{1}I+d_{2}U_{\nu}+d_{3}U_{\nu}^{2}%
\end{equation}%
\begin{equation}
U_{\nu}=V_{\nu}D_{j,m}V_{\nu}^{\dag}\ \cdot
\end{equation}
In fact, $U_{\nu}^{2}=1$ since the only acceptable solutions are for $m=2$ meaning that the coefficient $d_3$ does
not exist. This reduction in the degrees of freedom creates a degeneracy in the neutrino mass spectrum.
This way, for $D_{1,2}$ we get $m_{2}=m_{3}$ for $D_{2,2}$ we get $m_{1}=m_{3}$ and for
$D_{3,2}$ we get $m_{1}=m_{2}$ respectively. Thus, we can either have an exact discrete symmetry with
a degenerate neutrino spectrum or a non degenerate spectrum with a broken symmetry. In the latter case
one may choose a diagonal
neutrino mass matrix i.e. $M_d=\mathrm{Diagonal}[m_1,m_2,m_3]$ and construct the actual matrix through the relation
\[
M_{\nu}=V_{\nu}M_d V_{\nu}^{\dagger} \cdot
\]
The matrices $V_{l}$ and $V_{\nu}$ are arbitrary unitary matrices connected by
the relation
\begin{equation}
C=V_{l}^{\dagger}V_{\nu}\ .
\end{equation}
 We now want to generalise the TB mixing in order
to accomodate recent data. To that end, let us now assume that the mixing matrix
$C$ takes  the general form~(\ref{V123}).

We may again construct the matrix $\mathcal{T}$ as before, using the
new mixing matrix. As previously, we must require that
\begin{equation}
\operatorname{Im}\mathrm{Tr}\mathcal{T}=0,\ \operatorname{Re}\mathrm{Tr}%
\mathcal{T}=1+2\cos\frac{2\pi}{p}\ \cdot
\end{equation}

The initial search for viable models was done numerically. No solutions were found for $m \neq 2$.
Once a potentially working model is found, it can be elaborated analytically. In what follows, we
use the previous classification in terms of the diagonal matrices $D_{i,n},D_{j,m}$ and the three integers $n,m,p$ .

\section{Potentially viable Models}

\begin{enumerate}
\item Case $(1,1,3,2,p)$.  For this set of integers,  $\theta_{12}$  and $\theta_{13}$
are related as follows:
\[
\cos\theta_{12}=-\frac{2}{\sqrt{3}}\frac
{\cos\frac{\pi}{p}}{\cos\theta_{13}} \cdot
\]
  Computing  $\theta_{12}$ as a function of
 $p$, we find that the only acceptable value   is $p=3$. In this case, we obtain the following relations:
\[
\cos\theta_{12}   = -\frac{  2\cos\frac{\pi}{3}}%
{\sqrt{3}\cos\theta_{13}}\equiv-\frac{1}{\sqrt{3}\cos\theta_{13}}
\]
\[
\tan2\theta_{23}   =\frac{1}{2}\left(  \frac{\sin\theta_{13}}{\tan\theta
_{12}}-\frac{\tan\theta_{12}}{\sin\theta_{13}}\right) \cdot
\]
 Plots of the angles $\sin^{2}\theta_{12},\sin^{2}\theta_{12}$ as a
function of $\theta_{13}$ are shown in fig 1. We observe that, although
$\theta_{12}$ lies within the acceptable range, this is not so for
$\theta_{23}$. Therefore, this model is rejected.

\begin{figure}[b]
\centering
\includegraphics[scale=0.7]{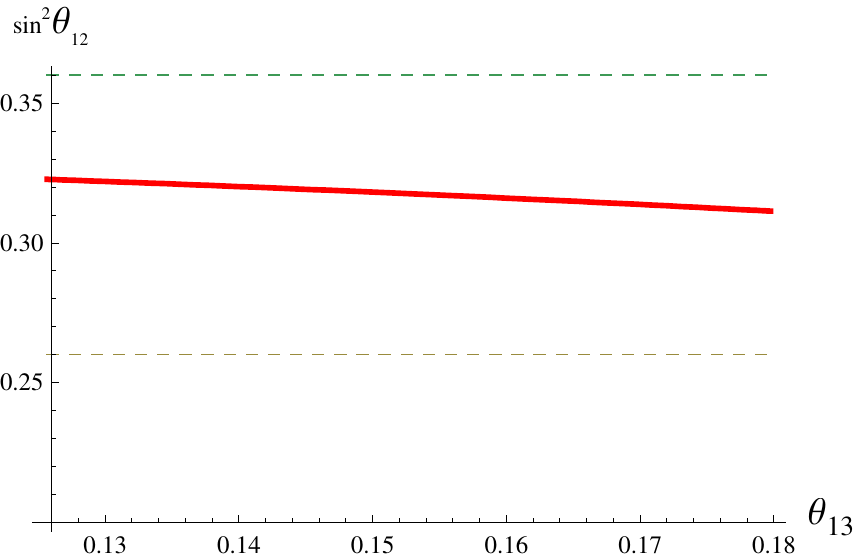}\;\includegraphics[scale=0.7]{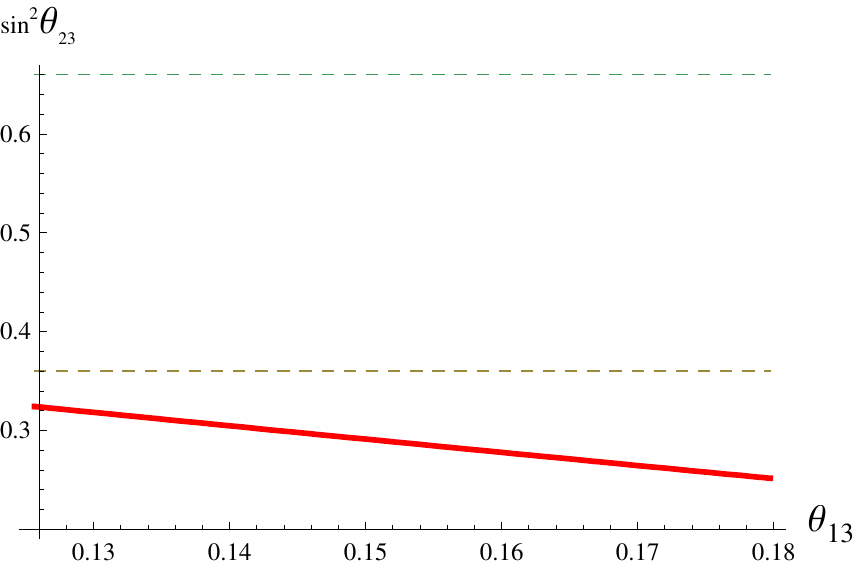}
\caption{Range for $\sin^{2}\theta_{12},\sin^{2}\theta_{23}$ as a function of
$\theta_{13}$ for the model $(2,2,3,3,3)$. We observe that $\sin^{2}\theta_{23}$
lies outside the $3\sigma$ range and therefore this model is incompatible with
the data.}%
\label{Painl}%
\end{figure}

\item Case $(2,2,3,2,p)$

Here, $\theta_{12}$ is given by
\[
\sin^{2}\theta_{12}=\frac{1}{6\cos^{2}\theta_{13}}\left(  1-2\cos\frac{2\pi
}{p}\right) \cdot
\]
The only acceptable value is $p=3$  giving
\[
\sin\theta_{12}\,\cos\theta_{13}=-\frac{1}{\sqrt{3}}%
\]
identical to $(1,1,3,2,p)$. However, the
angle $\theta_{23}$ differs
\[
\tan2\theta_{23}=-\frac{2\cot2\theta_{13}}{\sqrt{3-\sec^{2}\theta_{13}}} \cdot
\]

Plots for $\sin^{2}\theta_{12}$ and $\sin^{2}\theta_{23}$ vs $\theta_{13}$ are
given in figure \ref{model2}.  This time, both quantities lie within the
acceptable experimental ranges, hence this model is  compatible with the
data.
\begin{figure}[!b]
   \centering
   \includegraphics[scale=0.7]{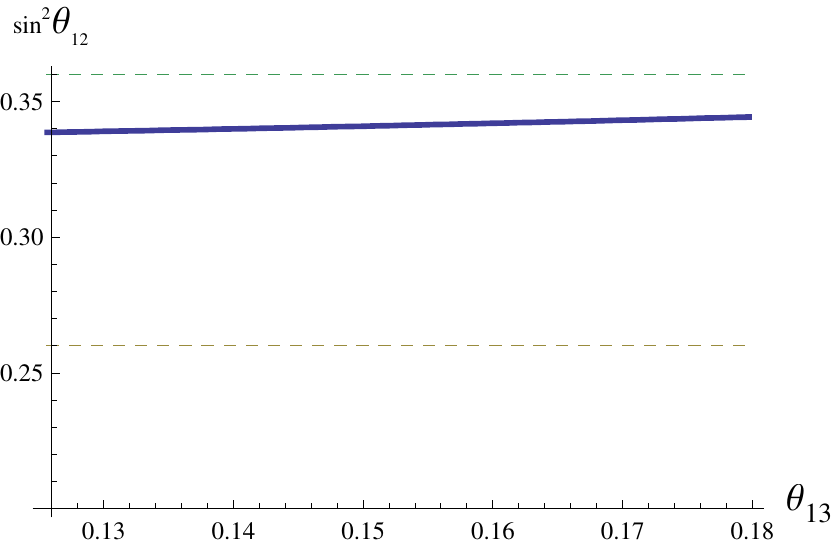}\;\includegraphics[scale=0.7]{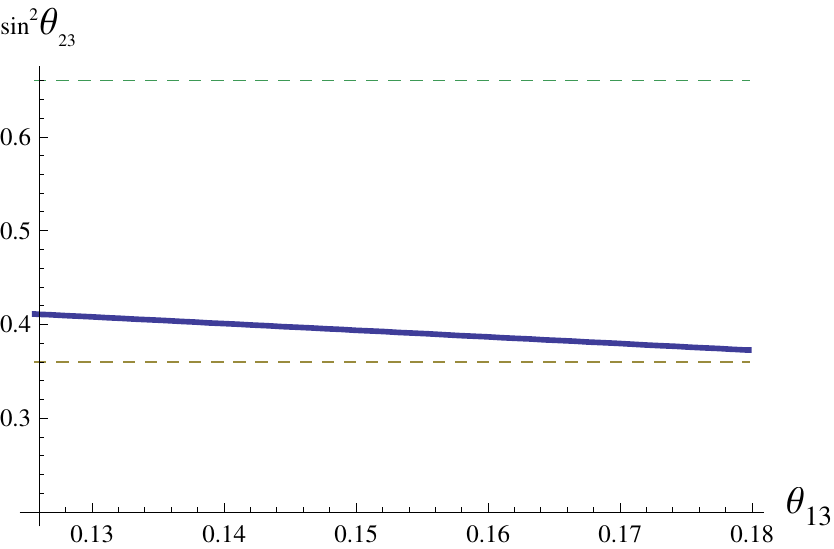}
   \caption{Range for $\theta_{12},\theta_{23}$ as a function of $\theta_{13}$
   for the  model  $(2,2,3,3,3)$.}
   \label{model2}
   \end{figure}

\item Case $(3,2,3,2,p)$

The resulting formula for $\theta_{12}$  is identical to the one in $(1,1,3,2,p)$
\[
\sin\theta_{12}=-\frac{2\cos\frac{\pi}{p}}{\sqrt{3}\cos\theta_{13}} \cdot
\]
Hence, as above, the only acceptable value is $p=3$, leading to the
constraint
\[
\sin\theta_{12}=-\frac{1}{\sqrt{3}\cos\theta_{13}}%
\]
as in $(1,1,3,2,p)$ and $(2,2,3,2,p)$. For $p=3$ the angle
$\theta_{23}$  is given by
\[
\tan2\theta_{23}=-\frac{2\cot2\theta_{13}}{\sqrt{3-\sec^{2}\theta_{13}}}\cdot
\]
Comparing with  previous cases, we observe that for $p=3$, as far as the
mixing is concerned, this model is identical to the second one, therefore it
is compatible with  data. Setting $\sin{\theta_{13}}=s$ the mixing matrix becomes

\[
C=\begin{bmatrix}
\sqrt{\frac{2}{3}-s^{2}} & -\frac{1}{\sqrt{3}} & s \\
\frac{\sqrt{3}}{2}s+\frac{1}{2}\sqrt{\frac{2}{3}-s^{2}} & \frac{1}{\sqrt{3}}
& \frac{1}{2}s-\frac{\sqrt{3}}{2}\sqrt{\frac{2}{3}-s^{2}} \\
-\frac{\sqrt{3}}{2}s+\frac{1}{2}\sqrt{\frac{2}{3}-s^{2}} & \frac{1}{\sqrt{3}}
& \frac{1}{2}s+\frac{\sqrt{3}}{2}\sqrt{\frac{2}{3}-s^{2}}%
\end{bmatrix} \cdot
\]

Since the only allowed generalization  requires that $n=3$, $m=2$, $p=3$ we are led to the
conclusion that the only symmetry group that can connect the charged lepton and the neutrino
 mass matrices is the discrete group $A_4$.  Observe that in the adopted formalism the
middle column of $C$ remains unchanged i.e. given by the column vector
$\{-\frac{1}{\sqrt{3}},\frac{1}{\sqrt{3}},\frac{1}{\sqrt{3}}\}^{T}$. It turns
out that the so constructed generalisation of the TB mixing matrix utilises
the freedom of making linear transformations inside the degenerate neutrino
subspace to create a non vanishing value for the $\theta_{13}$ angle. This
subspace is obviously orthogonal to the $\{-\frac{1}{\sqrt{3}},\frac{1}%
{\sqrt{3}},\frac{1}{\sqrt{3}}\}$ axis.

As an example, we compute the mixing matrix for $\theta_{13}=\pi/20$,
where the elements although exhibiting significant deviations from the TB case, are still consistent with data.

\[
V=\begin{bmatrix}
0.801371 & -0.57735 & 0.156434 \\
0.536162 & 0.57735 & -0.61579 \\
0.265209 & 0.57735 & 0.772225%
\end{bmatrix} \cdot
\]

\end{enumerate}

\section{The mass spectrum}

We have found two models with identical predictions with respect to
the mixing. The models differ only in the structure of the charged
lepton mass matrix implying different coefficients $c_{1,2,3}$ for
the two cases. Namely, if the mass matrix eigenvalues are given by
$m_{1}$,$ m_{2}$,$ m_{3}$, and we define $\Delta m_{ij}=m_i-m_j$,
the coefficient functions are given by

\be
\begin{split}
c_{1}^{2}&=-\frac{1}{8}\csc^{2}\frac{\pi}{n}\sec\frac{\pi}{n}\left[-2m_{2}\cos
\frac{\pi}{n} +m_3\, e^{\frac{3i\pi}{n}}+m_1\,e^{-\frac{3i\pi}{n}}
\right]
\\
c_{2}^{2}&=\frac{1}{4}\csc^{2}\frac{\pi}{n}\left[  \cos\frac{2\pi}{n}%
(\Delta m_{12}-\Delta m_{23})-i\sin\frac{2\pi}{n}\Delta m_{13}\right]
\\
c_{3}^{2}&=\frac{1}{8}\csc\frac{\pi}{n}\left[
\csc\frac{\pi}{n}(\Delta m_{23}- \Delta m_{12}
)+i\sec\frac{\pi}{n}\Delta m_{13}\right]
\end{split}
\ee
for  $D_{2,n}$ and
\be
\begin{split}
c_{1}^{3}&=-\frac{1}{8}\csc^{2}\frac{\pi}{n}\sec\frac{\pi}{n}\left[
-2m_{3} \cos\frac{\pi}{n}+m_2\,
e^{\frac{3i\pi}{n}}+m_1\,e^{-\frac{3i\pi}{n}}\right]
\\
c_{2}^{3}&=\frac{1}{4}\csc^{2}\frac{\pi}{n}\left[  \cos\frac{2\pi}{n}%
(\Delta m_{13}+\Delta m_{23})-i\sin\frac{2\pi}{n}\Delta m_{12}\right]
\\
c_{3}^{3}&=\frac{1}{8}\csc\frac{\pi}{n}\left[
-\csc\frac{\pi}{n}(\Delta m_{13}+\Delta m_{23})
+i\sec\frac{\pi}{n}\Delta m_{12}\right]  \ .
\end{split}
\ee
for  $D_{3,n}$.

The results found show that for leptons $n=3$ while for neutrinos we
get $m=2$. The corresponding coefficient functions are:
\begin{itemize}
\item Leptons
\end{itemize}
\be
\begin{split}
c_{1}^{2}&=\frac{1}{3}(m_{1}+m_{2}+m_{3})
\\
c_{2}^{2}&=\frac{1}{6}\left(  2m_{2}-m_{1}-m_{3}\right)  -\frac{i}{2\sqrt{3}%
}(m_{1}-m_{3})
\\
c_{3}^{2}&=\frac{1}{6}\left(  2m_{2}-m_{1}-m_{3}\right)  +\frac{i}{2\sqrt{3}%
}(m_{1}-m_{3})
\end{split}
\ee
and
\begin{equation}
\begin{split}
c_{1}^{3}&=\frac{1}{3}(m_{1}+m_{2}+m_{3})
\\
c_{2}^{3}&=\frac{1}{6}\left(  2m_{3}-m_{1}-m_{2}\right)  -\frac{i}{2\sqrt{3}%
}(m_{1}-m_{2})
\\
c_{3}^{3}&=\frac{1}{6}\left(  2m_{3}-m_{1}-m_{2}\right)  +\frac{i}{2\sqrt{3}%
}(m_{1}-m_{2})
\end{split}
\end{equation}

\begin{itemize}
\item Neutrinos
\end{itemize}
For the neutrinos $m=2$. The $D_2$ and the corresponding neutrino
matrices are given by:
\begin{equation}
\begin{split}
D_{2}&=\mathrm{Diagonal}\left[  -1,1,-1\right]
\\
M&=\mathrm{Diagonal}\left[  m_{1},m_{2},m_{3}\right]  \ \cdot
\end{split}
\end{equation}

In this case, it is clear that the neutrino mass spectrum turns out
to be degenerate since  ${D_{2}}^2=1$ implying $m_{1}=m_{3}$. In order to
establish a breaking pattern we must write
\[
M=d_{1}I+d_{2}D_{2}+R_{2}%
\]
where $R_{2}$ is the remainder term to be determined.
\[
R_{2}=%
\begin{bmatrix}
m_{1}-d_{1}+d_{2} & 0 & 0\\
0 & m_{1}-d_{1}-d_{2} & 0\\
0 & 0 & m_{3}-d_{1}+d_{2}%
\end{bmatrix}
\ \cdot
\]
Each non vanishing element must be proportional to the same mass difference in
order to have a breaking pattern. Since
\[
\left(  R_{1}\right)  _{11}-\left(  R_{1}\right)  _{33}=m_{1}-m_{3}%
\]
this mass difference is $m_{1}-m_{3}$ if no special relations between neutrino
masses are assumed. So we have
\begin{align*}
m_{1}-d_{1}+d_{2} &  =r_{1}\left(  m_{1}-m_{3}\right)  \\
m_{2}-d_{1}-d_{2} &  =r_{2}\left(  m_{1}-m_{3}\right)  \\
m_{3}-d_{1}+d_{2} &  =r_{3}\left(  m_{1}-m_{3}\right)
\end{align*}
with $r_{1}-r_{3}=1$. Solving for $d_1,d_2$ we get
\begin{align*}
d_{1} &  =\frac{1}{2}\left(  m_{1}+m_{2}\right)  -\frac{1}{2}\left(
r_{1}+r_{2}\right)  \left(  m_{1}-m_{3}\right)  \\
d_{2} &  =-\frac{1}{2}\left(  m_{1}-m_{2}\right)  +\frac{1}{2}\left(
r_{1}-r_{2}\right)  \left(  m_{1}-m_{3}\right)
\end{align*}
for arbitrary $r_1$ and $r_2$.
A different breaking pattern would require invariant relations
between the neutrino masses. For instance if we require that
\begin{align*}
m_{1}-d_{1}+d_{2} &  =r_{1}\left(  m_{1}-m_{2}\right)  \\
m_{2}-d_{1}-d_{2} &  =r_{2}\left(  m_{1}-m_{2}\right)  \\
m_{3}-d_{1}+d_{2} &  =r_{3}\left(  m_{1}-m_{2}\right)
\end{align*}
consistency implies that
\[
r_{3}=r_{1}+\frac{m_{3}-m_{1}}{m_{1}-m_{2}}%
\]
independent of the masses i.e. $\frac{m_{3}-m_{1}}{m_{1}-m_{2}}=\mu$, and
\[
d_{1}=\frac{1}{2}\left(  m_{1}+m_{2}\right)  -\frac{1}{2}\left(  r_{1}%
+r_{2}\right)  \left(  m_{1}-m_{2}\right)
\]%
\[
d_{2}=  \frac{1}{2}\left(  r_{1}%
-r_{2}-1 \right)  \left(  m_{1}-m_{2}\right) \cdot
\]

Finally, we end up our analysis by presenting the first order corrections
to the charged leptons mass matrix written as an expansion in terms of $s=\sin\theta_{13}$.
\[
m_{\ell}\approx A+sB
\]
where
\[
A=\left(
\begin{array}{ccc}
 \frac{1}{6} (4 {m_1}+{m_2}+{m_3}) & \frac{-2
   {m_1}+{m_2}+{m_3}}{3 \sqrt{2}} &
   \frac{{m_3}-{m_2}}{2 \sqrt{3}} \\
 \frac{-2 {m_1}+{m_2}+{m_3}}{3 \sqrt{2}} & \frac{1}{3}
   ({m_1}+{m_2}+{m_3}) & \frac{{m_3}-{m_2}}{\sqrt{6}}
   \\
 \frac{{m_3}-{m_2}}{2 \sqrt{3}} &
   \frac{{m_3}-{m_2}}{\sqrt{6}} & \frac{{m_2}+{m_3}}{2}
\end{array}
\right)
\]
and
\[
B=\left(
\begin{array}{ccc}
 \frac{({m_2}-{m_3}) }{\sqrt{2}} & \frac{1}{2}
   ({m_2}-{m_3})  & \frac{(2{m_1}-{m_2}-{m_3})
  }{\sqrt{6}} \\
 \frac{1}{2} ({m_2}-{m_3})  & 0 & \frac{(-2
   {m_1}+{m_2}+{m_3}) }{2 \sqrt{3}} \\
 \frac{(2 {m_1}-{m_2}-{m_3}) }{\sqrt{6}} & \frac{(-2
   {m_1}+{m_2}+{m_3}) }{2 \sqrt{3}} &
   \frac{({m_3}-{m_2}) }{\sqrt{2}}
\end{array}
\right)\cdot
\]

\section{Conclusions}

In this letter we have examined
the structure of the lepton and neutrino mass matrices  assuming that they
can be expanded in a basis of  finite group generators.
This  procedure has been proven useful for putting some order
on the enormous number of possibilities given in the literature in a
 systematic and mathematically consistent way.  Various models are
 classified by means of three integer numbers which define the group.
 The calculations show that only the $A_4$ group can reproduce current
 data by allowing a non zero value for the $\theta_{13}$  mixing angle.
 Exact group symmetry introduces a degeneracy in the neutrino spectrum which
  has to lifted by means of an external breaking mechanism.

  \vfill

{\bf{Aknowledgments}}. The authors would like to thank the CERN
theory group for its kind hospitality while this work was in
preparation. The research Project  is co-financed by the European
Union - European Social Fund (ESF) and National Sources, in the
framework of the program ``THALIS" of the ``Operational Program
Education and Lifelong Learning" of the National Strategic Reference
Framework (NSRF) 2007-2013.

\end{document}